\documentclass{article}

\usepackage[english]{babel}

\usepackage[letterpaper,top=2cm,bottom=2cm,left=3cm,right=3cm,marginparwidth=1.75cm]{geometry}

\usepackage{amsmath}
\usepackage{graphicx}
\usepackage[colorlinks=true, allcolors=blue]{hyperref}
\usepackage[round]{natbib}
\usepackage{makecell}
\usepackage{tabularx}
\usepackage{multirow}
\usepackage{booktabs}
\usepackage{float}
\usepackage{authblk}

\title{BayesPPDSurv: An R Package for Bayesian Sample Size Determination Using the Power and Normalized Power Prior for Time-To-Event Data}

\author[1]{Yueqi Shen\thanks{ys137@live.unc.edu}}
\author[2]{Matthew A. Psioda}
\author[1]{Joseph G. Ibrahim}
\affil[1]{Department of Biostatistics, University of North Carolina at Chapel Hill}
\affil[2]{GSK}

\begin{document}
\maketitle

\begin{abstract}
The \textbf{BayesPPDSurv} (Bayesian Power Prior Design for Survival Data) R package supports Bayesian power and type I error calculations and model fitting using the power and normalized power priors incorporating historical data with for the analysis of time-to-event outcomes. The package implements the stratified proportional hazards regression model with piecewise constant hazard within each stratum. The package allows the historical data to inform the treatment effect parameter, parameter effects for other covariates in the regression model, as well as the baseline hazard parameters. The use of multiple historical datasets is supported. A novel algorithm is developed for computationally efficient use of the normalized power prior. In addition, the package supports the use of arbitrary sampling priors for computing Bayesian power and type I error rates, and has built-in features that semi-automatically generate sampling priors from the historical data. We demonstrate the use of \textbf{BayesPPDSurv} in a comprehensive case study for a melanoma clinical trial design.
\end{abstract}

\section{Introduction: \textbf{BayesPPDSurv}} \label{sec:intro}

The incorporation of historical information in clinical trial design has become increasingly popular due to its potential of making trials more efficient. If the historical trial is sufficiently similar to the current trial, one can achieve more accurate point estimates and increased power \citep{Viele_2014}. One natural way of integrating historical information is through informative priors in a Bayesian framework. The power prior \citep{ibrahim_2000} is a popular class of informative priors that allow the incorporation of historical data through a discounted likelihood. It is constructed by raising the historical data likelihood to a power $a_0$, where $0 \le a_0 \le 1$. The discounting parameter $a_0$ can be fixed or modeled as random. When it is modeled as random and estimated jointly with other parameters of interest (denoted by $\theta$), the normalized power prior \citep{duan_2006} is recommended, as normalization is critical to enabling the prior to factor into a conditional distribution of $\theta$ given $a_0$ and a marginal distribution of $a_0$. 

The power prior is widely used due to its easy construction and intuitive interpretation \citep{ibrahim_2015}. The theoretical properties of the power prior and the normalized power prior have been extensively studied. For example, \cite{ibrahim_2003} show that the power prior is an optimal class of informative priors in the sense that it minimizes a convex sum of the Kullback–Leibler (KL) divergences between two posterior densities, in which one density is based on no incorporation of historical data, and the other density is based on pooling the historical and current data. \cite{shen_2023} show that the marginal posterior for the discounting parameter converges
to a point mass at zero if there is any discrepancy between the historical and current data. They also show that the marginal posterior for $a_0$ does not converge to a point mass at one when the datasets are fully compatible, and yet, for an i.i.d. normal model and finite sample size, the marginal posterior for $a_0$ always has most mass around one when the datasets are fully compatible. \cite{chen_2006} and \cite{shen_bhm} establish the analytic connection between the the power and the normalized power priors and Bayesian hierarchical models. The normalizing constant in the normalized power prior is analytically intractable when there are covariates, except in the case of the normal linear model.  \cite{Carvalho_Ibrahim_2021} propose a bisection-type algorithm for computing the normalizing constant. In our package, \textbf{BayesPPDSurv} (Bayesian Power Prior Design for Survival Data), we propose a novel algorithm that provides an arbitrarily accurate approximation to the normalized power prior for $\theta$ itself, avoiding the need to compute the normalizing constant and thus providing significant computational efficiency. 

The focus of the \textbf{BayesPPDSurv} package is on applying the power prior and normalized power prior to time-to-event outcomes for sample size determination and/or data analysis. In particular, we implement the proportional hazards model \citep{cox} with piecewise constant baseline hazard \citep{Friedman_1982}. Modeling the baseline hazard with a piecewise constant function is a widely used approach in Bayesian analysis \citep{ibrahim_2001}. Several \textbf{R} packages on CRAN implement the piecewise constant hazard model. The \textbf{pch} package \citep{pch} implements the piecewise constant hazard model for censored and truncated data. The \textbf{gsDesign} package \citep{gsDesign} supports the piecewise constant hazard model for group sequential clinical trial design. These packages do not allow the inclusion of historical data. %The \textbf{gsbDesign} package \citep{gsbDesign} implements group sequential clinical trial design that allows the integration of prior information for normal outcomes. 
There are several R packages that implement the power prior or variations of the power prior for the purpose of model estimation or design. The \textbf{BayesCTDesign} package \citep{bayesctdesign} supports two-arm randomized Bayesian trial design using historical control data with the power prior for a variety of outcome models, including the Weibull and piesewise constant hazard models for time-to-event outcomes. However, it does not allow using the historical data to inform the treatment effect parameter or parameter effects for other covariates. The \textbf{bayesDP} package \citep{bayesDP} implements the discounted power prior for single arm and two-arm clinical trials where the discounting parameter is determined by a discounting function estimated based on a measure of prior-data conflict. While it accommodates borrowing historical information for the treatment effect with the piecewise constant hazard model, it does not allow additional covariates, and it must be used in conjunction with the package \textbf{bayesCT} \citep{bayesCT} for trial design. There are three R packages that implement the normalized power prior where $a_0$ is modeled as random. The \textbf{NPP} package \citep{NPP} supports posterior sampling using the normalized power prior for Bernoulli, normal, multinomial and Poisson models, as well as for the normal linear model. The \textbf{hdbayes} \citep{hdbayes} package implements several methods that leverage historical data for generalized linear models, including the power prior and the normalized power prior. These two packages do not accommodate models for time-to-event outcomes, nor do they perform sample size determination. The R package \textbf{BayesPPD} \citep{BayesPPD} supports trial analysis and design using the power prior and the normalized power prior for generalized linear models, but it does not support models for time-to-event outcomes. 

Our package \textbf{BayesPPDSurv} \citep{BayesPPDSurv} addresses an important gap by providing a suite of functions for Bayesian power and type I error rate calculations and model fitting after incorporating historical data with the power prior and the normalized power prior for time-to-event data. It implements the stratified proportional hazards regression model with piecewise constant hazard within each stratum. The package allows the historical data to inform the treatment effect parameter, parameter effects for other covariates in the hazard ratio regression model, as well as the baseline hazard parameters. The time interval partition for the piecewise baseline hazards can be stratified. The discounting parameter $a_0$ can be fixed or modeled as random. The use of multiple historical datasets is supported. For sample size determination, we consider the simulation-based method developed in \cite{Chen_2011} utilizing the sampling and fitting priors \citep{Gelfand_2002} as applied in \cite{Psioda_Ibrahim_2019}. The package supports the use of arbitrary sampling priors for computing Bayesian power and type I error rates, and has built-in features that semi-automatically generate sampling priors from the historical data. \textbf{BayesPPDSurv} is computationally efficient. It implements the slice sampler \citep{slice} with Rcpp \citep{Rcpp}, and functions for analysis take less than a minute to execute in most instances.

The rest of the article proceeds as follows. In section 2, we describe the theoretical details of the methods implemented in the package. In section 3, we provide details on using the package and its various features. In section 4, we present a comprehensive case study for a melanoma clinical trial design with example code. The article is concluded with a brief discussion.

\section{Theoretical Framework}
\subsection{The Power Prior and the Normalized Power Prior}

Let $D$ denote data from the current study and $D_0$ denote data from a historical study. Let $\theta$ denote the model parameters and $L(\theta|D)$ denote a general likelihood function associated with a given outcome model, such as a generalized linear model (GLM) or a survival model. The power prior \citep{ibrahim_2000} is defined as $$\pi(\theta|D_0, a_0) \propto L(\theta|D_0)^{a_0}\pi_0(\theta),$$ where $0 \le a_0 \le 1$ is a discounting parameter for the historical data likelihood, and $\pi_0(\theta)$ is the initial prior for $\theta$. The parameter $a_0$ allows one to control the influence of the historical data on the posterior distribution. When $a_0=0$, historical information is disregarded and the power prior becomes equivalent to the initial prior $\pi_0(\theta)$. Conversely, when $a_0=1$, the power prior corresponds to the posterior distribution of $\theta$ given the historical data and the initial prior.    

The power prior can easily accommodate multiple historical datasets. Suppose there are $J$ historical datasets denoted by $D_{0j}$ for $j=1,\cdots, J$ and let $D_0=(D_{01}, \cdots, D_{0J})$. The power prior becomes $$\pi(\theta|D_0, a_0) \propto \prod_{j=1}^J L(\theta|D_{0j})^{a_{0j}}\pi_0(\theta),$$ where $a_0 = (a_{01},\cdots,a_{0J})'$ are dataset-specific discounting parameters and $0\le a_{0j} \le 1$ for $j=1,\cdots,J$.

Modeling $a_0$ as random allows one to represent uncertainty in how much the historical data should be discounted. \cite{duan_2006} introduce the \emph{normalized power prior}, given by $$\pi(\theta, a_0|D_0) = \pi(\theta|D_0, a_0)\pi(a_0) = \frac{L(\theta|D_0)^{a_0}\pi_0(\theta)}{c(a_0)}\pi(a_0),$$ where $\pi(a_0)$ is the initial prior for $a_0$. The normalized power prior specifies a conditional prior for $\theta$ given $a_0$ and a marginal prior for $a_0$. The normalizing constant, $$c(a_0)=\int L(\theta|D_0)^{a_0}\pi_0(\theta) d\theta,$$ is often analytically intractable and requires Monte Carlo methods for estimation. In  \textbf{BayesPPDSurv}, we implement a novel and computationally viable method for using the normalized power prior that avoids the need to compute the normalizing constant.

\subsection{The Piecewise Constant Hazard Proportional Hazards (PWCH-PH) Model}
In \textbf{BayesPPDSurv}, we implement the stratified proportional hazards model with piecewise
constant baseline hazard within each stratum, which is a common approach for Bayesian analysis of time-to-event data \citep{ibrahim_2001}. 

In our implementation, we allow the incorporation of a stratification variable with $S$ levels, where $s=1,\dots,S$ is the stratum index. Let $\lambda_s(t)$ denote the baseline hazard and $\Lambda_s(t)$ denote the cumulative baseline hazard for stratum $s$. Let $t_i$ denote the time to event and $c_i$ denote the time to censorship for subject $i$, $i=1,\dots,n$. Suppose $\nu_i=I[t_i \le c_i]$ denotes the indicator that an event is observed for subject $i$,  $y_i=\min(t_i, c_i)$ denotes the individual's observation time, and $x_i=(x_{i1},\dots,x_{iP})'$ denotes a $P \times 1$ vector of covariates, where without loss of generality we assume $x_{i1}$ is the treatment indicator. If there are no additional covariates, $x_i$ will consist of the the treatment indicator only. The current data consists of $D = \{(y_{i}, x_{i}, \nu_{i}), i=1,\cdots, n\}$. Further, $\beta=(\beta_1,\dots,\beta_P)'$ denotes a $P \times 1$ vector of regression coefficients, where $\beta_1$ is the coefficient for the treatment indicator, and $\lambda$ denotes the set of all baseline hazard parameters. Then, the likelihood of a stratified proportional hazards model is given by
\begin{align*}
L(\beta,\lambda|D)\propto &\prod_{i=1}^n\{\lambda_{si}(y_i)\exp(x_i'\beta)\}^{\nu_i}\exp\{-\Lambda_{si}(y_i)\exp(x_i'\beta)\}.
\end{align*}

The package allows the baseline hazard and time interval partition to vary across the $S$ levels of the stratification variable. For stratum $s$, we partition time into $K_s$ intervals, where $k=1,\dots,K_s$ is the interval index, with change points $0=t_{s,0} < t_{s,1} < \cdots < t_{s,K_s} = \infty$. Suppose $\lambda_{sk} > 0$ denotes the constant hazard over interval $I_{s,k}=(t_{s,k-1}, t_{s,k}]$ for stratum $s$, $\nu_{ik}$ denotes the indicator that the event occurred in interval $I_{s_i,k}$, $r_{ik}$ denotes the subject's time at risk in interval $I_{s_i,k}$, and $G_s$ denotes the set of indices corresponding to subjects from stratum $s$. 
%Then the likelihood is $$\prod_{i=1}^n\prod_{k=1}^K(\lambda_k\exp(x_i'\beta))^{\delta_{ik}\nu_i}\exp\{-\delta_{ik}[\lambda_k(y_i-t_{k-1})+\sum_{g=1}^{k-1}\lambda_g(t_g-t_{g-1})]\exp(x_i'\beta)\}.$$
%The log likelihood is $$\sum_{i=1}^n\sum_{k=1}^K\delta_{ik}\nu_i[\log(\lambda_k)+x_i'\beta]-\delta_{ik}[\lambda_k(y_i-t_{k-1})+\sum_{g=1}^{k-1}\lambda_g(t_g-t_{g-1})]\exp(x_i'\beta).$$ The historical log likelihood is $$\sum_{i=1}^{n_0}\sum_{k=1}^{K_0}\delta_{0ik}\nu_{0i}[\log(\lambda_k)+x_{0i}'\beta]-\delta_{0ik}[\lambda_k(y_{0i}-t_{k-1})+\sum_{g=1}^{k-1}\lambda_g(t_{g}-t_{g-1})]\exp(x_{0i}'\beta).$$ 
Then the likelihood of a PWCH-PH model can be re-written as
\begin{align*}
L(\beta,\lambda|D)\propto\prod_{s=1}^S\prod_{k=1}^{K_s}\lambda_{sk}^{\sum\limits_{i\in G_s}\nu_{ik}}\exp\left\{-\lambda_{sk}\left(\sum_{i\in G_s}\exp(x_i'\beta)r_{ik}\right)\right\}\times \prod_{i=1}^n\exp(x_i'\beta)^{\nu_i}.
\end{align*}
%The log likelihood is $$\sum_{s=1}^S\sum_{k=1}^{K_s}\left[\sum\limits_{i\in G_s}\nu_{ik}\log(\lambda_{sk})-\lambda_{sk}\left(\sum_{i\in G_s}\exp(x_i'\beta)r_{ik}\right)\right] + \sum_{i=1}^n \nu_i(x_i'\beta)$$

\subsection{Power Prior for the PWCH-PH Model}
Suppose there are $J$ historical datasets, $j=1,\dots,J$, each with sample size $n_{0j}$. Analogously, the $j^{th}$ historical dataset consists of $D_{0j} = \{(y_{0ji}, x_{0ji}, \nu_{0ji}), i=1,\cdots, n_{0j}\}$, where $y_{0ji}$, $x_{0ji}$ and $\nu_{0ji}$ correspond to the individual’s observation time, covariate vector and event indicator, respectively, for subject $i$ in historical data $j$. Suppose $G_{sj}$ denotes the set of indices corresponding to subjects from stratum $s$ in historical dataset $j$, the baseline hazard for the historical data for stratum $s$ and interval $k$ is denoted by $\lambda_{0sk}$, and $\lambda_0$ denotes the set of all baseline hazard parameters for the historical data. Let $a_0=(a_{01},\dots,a_{0J})$ denote the discounting parameters for the $J$ historical datasets, where $a_{0j}$ discounts the $j^{th}$ historical dataset. Then, the power prior for $\beta$ is given by
\begin{align*}
&\pi(\beta|D_0, \lambda_0, a_0) \\
\propto &\prod_{j=1}^JL(\beta|\lambda_0,D_{0j})^{a_{0j}}\times\pi_0(\beta)\\
\propto&\prod_{j=1}^J\prod_{s=1}^S\prod_{k=1}^{K_s}\lambda_{0sk}^{a_{0j}\left(\sum_{i\in G_{sj}}\nu_{ik}\right)}\exp\left\{-a_{0j}\lambda_{0sk}\left(\sum_{i\in G_{sj}}\exp(x_{0i}'\beta)r_{ik}\right)\right\}\times\prod_{j=1}^J\prod_{i=1}^{n_{0j}}\exp(x_{0i}'\beta)^{\nu_{0i}a_{0j}}\times\pi_0(\beta).
%\propto&\prod_{s=1}^S\prod_{k=1}^{K_s}\lambda_{0sk}^{\sum_{j=1}^Ja_{0j}\left(\sum_{i\in G_{sj}}\nu_{ik}\right)}\exp\left\{-\sum_{j=1}^Ja_{0j}\lambda_{0sk}\left(\sum_{i\in G_{sj}}\exp(x_i'\beta)r_{ik}\right)\right\}\times \prod_{j=1}^J\prod_{i=1}^{n_{0j}}\exp(x_i'\beta)^{\nu_ia_{0j}}\times\pi_0(\beta)
\end{align*}
In this formulation, the current and historical datasets have different baseline hazard parameters ($\lambda$ and $\lambda_0$), so the historical data does not directly inform the inference for $\lambda$. If we assume the current and historical data share the same set of baseline hazard parameters, then $\lambda_0=\lambda$, and the historical information is used to estimate $\lambda$ directly.

For the initial prior on $\beta$, the package allows several choices of priors, including the uniform improper prior and independent normal priors for each component of $\beta$. For the priors on the $\lambda_{sk}$'s, the package allows independent Gamma priors, independent normal priors on $\log(\lambda_{sk})$, and the improper prior $\pi(\lambda) \sim \prod_{s=1}^S \prod_{k=1}^K \lambda_{sk}^{-1}.$ If the baseline hazard parameters are not shared between the current and historical data, the prior choices supported for $\lambda_0$ are independent Gamma priors, independent normal priors on $\log(\lambda_{0sk})$, and the improper prior $\pi(\lambda_0) \sim \prod_{s=1}^S \prod_{k=1}^K \lambda_{0sk}^{-1}$.

\subsection{Implementing the Normalized Power Prior for the PWCH-PH Model}

\textbf{BayesPPDSurv} supports the use of a normalized power prior for the case where $a_0$ is modeled as random and the baseline hazard parameters are \emph{unshared} between the current and historical data (i.e., $\lambda \neq \lambda_0$). We use independent normal initial priors for $\beta$, $N(\mu_p, \sigma_p^2)$ for each $\beta_p$,  independent gamma priors for $\lambda$ and $\lambda_0$ and independent beta priors for $a_0$. Specifically, we use $$\lambda_{sk} \sim \text{Gamma}(a_{sk}, b_{sk}),$$ $$\lambda_{0sk} \sim \text{Gamma}(c_{sk}, d_{sk}),$$ and $$a_{0j} \sim beta(u_{j}, v_{j}),$$ where $a_{sk}$, $b_{sk}$, $c_{sk}$, $d_{sk}$, $u_{j}$ and $v_{j}$ are user-specified hyperparameters.
The joint prior for $\beta$, $\lambda$, $\lambda_0$ and $a_0$ can now be written as $$\pi(\beta,\lambda,\lambda_0,a_0|D_0)\propto\pi(\lambda_0|\beta,a_0,D_0) \times \pi(\beta|D_0, a_0) \times \pi(a_0) \times \pi(\lambda).$$ 
We first factor $\pi(\beta,\lambda_0|D_0,a_0)$ into $\pi(\beta|D_0, a_0) \times \pi(\lambda_0|\beta,a_0,D_0)$.
We have 
\begin{align*}
&\pi(\beta,\lambda_0|D_0,a_0)\propto\prod_{j=1}^JL(\beta, \lambda_0|D_{0j})^{a_{0j}}\pi_0(\beta)\pi(\lambda_0)\\
=&\prod_{s=1}^S\prod_{k=1}^{K_s}\lambda_{0sk}^{\sum_{j=1}^Ja_{0j}\left(\sum_{i\in G_{sj}}\nu_{ik}\right)}\exp\left\{-\sum_{j=1}^Ja_{0j}\lambda_{0sk}\left(\sum_{i\in G_{sj}}\exp(x_i'\beta)r_{ik}\right)\right\}\frac{d_{sk}^{c_{sk}}}{\Gamma(c_{sk})}\lambda_{0sk}^{c_{sk}-1}\exp(-d_{sk}\lambda_{0sk})\\
&\times \prod_{p=1}^P N(\mu_p,\sigma_p^2)\times\prod_{j=1}^J\prod_{i=1}^{n_{0j}}\exp(x_i'\beta)^{\nu_ia_{0j}} \\
=&\prod_{p=1}^P N(\mu_p,\sigma_p^2)\prod_{j=1}^J\prod_{i=1}^{n_{0j}}\exp(x_i'\beta)^{\nu_ia_{0j}}\prod_{s=1}^S\prod_{k=1}^{K_s}\frac{\frac{d_{sk}^{c_{sk}}}{\Gamma(c_{sk})}\Gamma(p_{sk})}{q_{sk}^{p_{sk}}} \times \text{Gamma}(\lambda_0; p_{sk}, q_{sk}),
\end{align*}
where $p_{sk}=\sum_{j=1}^Ja_{0j}\left(\sum_{i\in G_{sj}}\nu_{ik}\right)+c_{sk}$ and $q_{sk}=\sum_{j=1}^Ja_{0j}\left(\sum_{i\in G_{sj}}\exp(x_i'\beta)r_{ik}\right)+d_{sk}$.\\
Then $$\pi(\lambda_{0sk}|\beta,a_0,D_0)=\text{Gamma}(p_{sk}, q_{sk}),$$ and
\begin{equation}\label{kernel}
\pi(\beta|D_0, a_0) \propto \prod_{s=1}^S\prod_{k=1}^{K_s}q_{sk}^{-p_{sk}}\prod_{j=1}^J\prod_{i=1}^{n_{0j}}\exp(x_i'\beta)^{\nu_ia_{0j}}\prod_{p=1}^P N(\mu_p,\sigma_p^2).
\end{equation}
 When using the normalized power prior in \textbf{BayesPPDSurv}, for simplicity of computation, we focus on inferences for $\beta$ and $\lambda$ only, so we marginalize over $\lambda_0$ and $a_0$ to obtain 
\begin{align*}
\pi(\beta,\lambda|D_0) &= \int\int\pi(\lambda_0|\beta,a_0,D_0) \times \pi(\beta|D_0, a_0) \times \pi(a_0) \times \pi(\lambda)d\lambda_0da_0\\
&=\int\pi(\beta|D_0, a_0)\pi(a_0)da_0 \times \pi(\lambda).
\end{align*}
The normalizing constant for equation \eqref{kernel} is analytically intractable. In \textbf{BayesPPD}, we implemented an algorithm \citep{shen_RJ} that approximates the normalizing constant with the partition weighted kernel estimator \citep{pwk_2018}. In \textbf{BayesPPDSurv}, instead of approximating the normalizing constant, we approximate the normalized power prior for $\beta$ itself, $\pi(\beta|D_0)=\int\pi(\beta|D_0, a_0)\pi(a_0)da_0$, with a multivariate normal distribution, utilizing the Bayesian Central Limit Theorem \citep{chen_1985}. We propose a novel algorithm to approximate $\pi(\beta|D_0)$ as follows: 
\begin{enumerate}
\item  Sample $a_{01}, \dots, a_{0L}$ from $\pi(a_0)$, where $L=10,000$, for example. Note that each $a_{0l}$ is a $J \times 1$ vector corresponding to the $J$ historical datasets. 
\item Given each $a_{0l}$, use slice sampling \citep{slice} to obtain a sample of $\beta_l$ (after burn-in) based on the kernel of $\pi(\beta|a_{0l}, D_0)$ in equation \eqref{kernel}.
\item Given $\beta_1, \dots, \beta_L$, the distribution $\pi(\beta|D_0)$ can be approximated by a multivariate normal distribution or a mixture of multivariate normals with the latter leading to an arbitrarily accurate approximation. To approximate with a single multivariate normal, one can compute the mean $\bar{\beta}$ and covariance $\bar{\Sigma}$ of $\beta_1, \dots, \beta_L$, and approximate $\pi(\beta|D_0)$ with $N_P(\bar{\beta}, \bar{\Sigma})$. 
\end{enumerate}
By default, the function \textit{phm.random.a0()} approximates $\pi(\beta|D_0)$ with a single multivariate normal distribution, $N_P(\bar{\beta}, \bar{\Sigma})$. The package allows the user to approximate $\pi(\beta|D_0)$ with any finite mixture of multivariate normal distributions as well. To achieve this, the user can take the output of the function \textit{approximate.prior.beta()}, which are discrete samples of $\beta_l$, $l=1,\dots,L$, and use external software to compute a mixture of multivariate normal distributions that best approximates the samples. Further details are given in section \ref{usage}. In Table \ref{power_npp} of section \ref{melanoma}, we show that this method provides highly accurate approximations. 

%%%%%%%%%%%%%%%%%%%%%%%%%%%%%%%%%%%%%%%%%%

\subsection{Bayesian Sample Size Determination}

To perform Bayesian sample size determination, we use the simulation-based procedure proposed in \cite{Chen_2011}. We assume the first column of the covariate matrix is the treatment indicator,
and the corresponding parameter is $\beta_1$, the log hazard ratio. 
By default, the hypotheses are given by
$$H_0: \beta_1 \ge \delta$$ and $$H_1: \beta_1 < \delta,$$
where $\delta$ is a prespecified value. To test hypotheses in the opposite direction, i.e., $H_0: \beta_1 \le \delta$ and $H_1: \beta_1 < \delta$, one can set the parameter \textit{nullspace.ineq} to "$<$".

Now we discuss the simulation-based procedure used to estimate the Bayesian type I error rate and power. Let $\Theta_0$ and $\Theta_1$ denote the parameter spaces corresponding to $H_0$ and $H_1$. Let $y^{(n)}$ denote the simulated current data associated with a sample size of $n$, $\theta=(\beta, \lambda, \lambda_0)$ denotes the model parameters, $\pi^{(s)}(\theta)$ denotes the sampling prior, and $\pi^{(f)}(\theta)$ denotes the fitting prior. The sampling prior is used to generate the hypothetical data while the fitting prior is used to fit the model after the data are generated. Let $\pi_0^{(s)}(\theta)$ denote the sampling prior that only puts mass in the null region, i.e., $\theta \subset \Theta_0$, and $\pi_1^{(s)}(\theta)$ denotes the sampling prior that only puts mass in the alternative region, i.e., $\theta \subset \Theta_1$. Let $q_0^{(n)}$ and $q_1^{(n)}$ denote the Bayesian type I error rate and power, respectively, associated with a sample size of $n$. Let $B$ denote the number of simulated trials. To compute the 
 Bayesian power or type I error, the following algorithm is used for each simulated trial $b$:
\begin{itemize}
\item{Step 1: }{Generate $\theta^{(b)} \sim \pi_j^{(s)}(\theta)$, $j=0,1$. The choice of sampling priors is described in section \ref{sampling_priors}.}
\item{Step 2: }{Generate $y^{(b)} \sim f(y^{(b)}|\theta^{(b)})$. Data generation for the PWCH-PH model is detailed in section \ref{data_sim}.}
\item{Step 3: }{Estimate the posterior distribution $\pi(\theta|y^{(b)}, D_0, a_0)$ and the posterior probability $P(\beta_1 < \delta|y^{(b)}, \pi^{(f)},D_0, a_0)$. Model fitting is performed using a custom slice sampler written in C++ via Rcpp \citep{slice,Rcpp}}.
\item{Step 4: }{Compute the indicator $r^{(b)}=I\{P(\beta_1 < \delta|y^{(b)}, \pi^{(f)}, D_0, a_0) \ge \gamma$\}, where the constant $\gamma > 0$ is a prespecified posterior probability threshold for rejecting the null hypothesis (e.g., $0.975$).}
\end{itemize}

The estimate of $q_j^{(n)}$ is given by $\frac{1}{B}\sum_{b=1}^B r^{(b)}$, where $j=0,1$. The quantity $q_{0}^{(n)}$ corresponding to $\pi^{(s)}(\theta)=\pi_0^{(s)}(\theta)$ is the Bayesian type I error rate, while $q_{1}^{(n)}$ corresponding to $\pi^{(s)}(\theta)=\pi_1^{(s)}(\theta)$ is the Bayesian power.
To identify the desired sample size, we can compute $n_{\alpha_0} = \min\{n: q_{0}^{(n)} \le \alpha_0\}$ and $n_{\alpha_1} = \min\{n: q_{1}^{(n)} \ge 1-\alpha_1\}$ for given $\alpha_0 > 0$ and $\alpha_1 > 0$, and therefore, the sample size is taken to be max$\{n_{\alpha_0}, n_{\alpha_1}\}$. Common choices of $\alpha_0$ and $\alpha_1$ include $\alpha_0=0.05$ and $\alpha_1=0.2$. These choices guarantee that the Bayesian type I error rate is at most $0.05$ and the Bayesian power is at least $0.8$.

\subsection{Data Simulation for the PWCH-PH Model }\label{data_sim}
Following \cite{psioda_adapt}, we describe the steps for simulating the observed data for the PWCH-PH model. We simulate the complete data for subject $i$ through the following procedure:
\begin{enumerate}
    \item Simulate the enrollment time $r_i$. The package allows choices of the uniform or the exponential distribution for the distribution for enrollment times.
    \item Simulate the treatment indicator $x_{i1}$ from a Bernoulli distribution with user-specified randomization probability. The default value is $0.5$.
    \item Simulate the additional covariates $x_{i2},\dots,x_{iP}$. The package samples one row of covariates with replacement from the combined covariate matrix of all historical datasets. 
    \item Simulate the stratum index $s_i$. The package samples with replacement from the combined stratum vector of all historical datasets. 
    \item Compute $\phi_i=\exp(x_i'\beta)$ and $\theta_{i,k}=\lambda_{s_i,k}\phi_i$ for $k=1,\dots,K_{s_i}$.
    \item Simulate the event time $t_i$ as follows:
    \begin{enumerate}
        \item Set $k=1$.
        \item Simulate $t_i \sim \text{Exponential}(\text{rate}=\theta_{i,k}) + t_{s_i, k-1}$.
        \item If $t_i > t_{s_i,k}$, then increment $k$ by one and return to step (b), otherwise terminate. 
    \end{enumerate}
    \item Simulate the censorship time $c_i$. Choices for the distribution of censorship times include the uniform distribution, the exponential distribution, or a user-specified constant time.
    \item Compute the observation time $\tilde{y}_i=\min(t_i,c_i)$ and the event indicator $\tilde{\nu}_i=I(t_i \le c_i)$.
    \item Compute the elapsed time $e_i=r_i+y_i$.
\end{enumerate}
The above procedure yields a hypothetical complete dataset corresponding to a scenario where all subjects are followed until the event is observed or they drop out. One constructs the observed dataset from the complete dataset as follows:  
\begin{enumerate}
    \item Determine the time $T$ of target event number $\nu$ in the complete dataset.
    \item Remove subjects with $r_i \ge T$.
    \item For the remaining subjects, 
    \begin{enumerate}
        \item If $e_i > T$, set $y_i=T-r_i$ and $\nu_i=0$.
        \item If $e_i \le T$, set $y_i=\tilde{y}_i$ and $\nu_i=\tilde{\nu}_i$.
    \end{enumerate}
\end{enumerate}
In the above, the quantity $\nu$ represents the target number of events that is desired to trigger the primary analysis for the simulated trial. The user can also determine the minimum amount of time ($T_{\min}$) and maximum amount of time ($T_{\max}$) that subjects are followed. If $T < T_{\min}$, then $T$ is replaced by $T_{\min}$. Administrative censoring occurs at $T_{\max}$.

\section{Using \textbf{BayesPPDSurv}}\label{usage}
\textbf{BayesPPDSurv} supports the incorporation of multiple historical datasets using the power prior with fixed $a_0$ or using the normalized power prior with $a_0$ modeled as random. It contains two categories of functions, functions for model fitting ("\textit{phm.fixed.a0}" and "\textit{phm.random.a0}") and functions for sample size determination ("\textit{power.phm.fixed.a0}" and "\textit{power.phm.random.a0}"). Functions for model fitting return posterior samples of the parameters, while functions for sample size determination return estimates of the Bayesian power or type I error rate. 

When $a_0$ is modeled as random, the function \textit{approximate.prior.beta()} performs slice sampling to generate samples of $\beta$ from the normalized power prior. By default, the function \textit{phm.random.a0()} approximates the normalized power prior for $\beta$, $\pi(\beta|D_0)$, with a single multivariate normal distribution. To accommodate greater accuracy of approximation, the package allows the user to approximate $\pi(\beta|D_0)$ with a finite mixture of multivariate normal distributions as well. Specifically, the user can take the output of the \textit{approximate.prior.beta()} function, which is a matrix of samples of $\beta$ from the normalized power prior, and use external software to compute a mixture of multivariate normal distributions that best approximates the samples. The resulting mixture distribution can then be the input to the \textit{prior.beta.mvn} argument of \textit{phm.random.a0()}, which is a list of lists, where each list has three elements, consisting of the mean vector, the covariance matrix and the weight of each multivariate normal distribution. Then, $\pi(\beta|D_{0})$ is approximated by the mixture of the multivariate normal distributions provided. Since we use $\lambda_0$ and $a_0$ as auxiliary variables for the approximation for $\pi(\beta|D_0)$, the user specifies the hyperparameters for the priors for $\lambda_0$ and $a_0$ in the function \textit{approximate.prior.beta()}. The hyperparameters for the priors for $\lambda$ and the initial prior for $\beta$ are specified in the function \textit{phm.random.a0()}. 

The package allows the time interval partition to vary across the $S$ levels of the stratification variable. The user must specify the total number of intervals for each stratum, \textit{n.intervals}. Then for each stratum, by default, the change points are assigned so that an approximately equal number of events are observed in all the intervals for the pooled historical and current datasets. The user can also specify the change points for each stratum. When $a_0$ is fixed, by default, the baseline hazard parameters are unshared (i.e., $\lambda \neq \lambda_0$) between the current and historical data. If \textit{shared.blh=TRUE}, baseline hazard parameters are shared and historical information is used to estimate the baseline hazard parameters. When $a_0$ is modeled as random, the package only supports unshared baseline hazards. 

For study design applications, the user can manipulate many attributes of the data generation process, including the enrollment time distribution (uniform or exponential), the randomization probability for the treated group, the censorship time distribution (uniform, exponential or constant), the probability of subjects dropping out of the study (non-administrative censoring), the dropout time distribution (uniform), and the minimum and maximum amount of time that subjects are followed.

\subsection{Sampling Priors}\label{sampling_priors}
Our implementation in \textbf{BayesPPDSurv} does not assume any particular distribution for the sampling
priors. The user specifies discrete approximations of the sampling priors by providing a matrix or list of parameter values and the algorithm samples with replacement from the matrix or the list as the first step of the data generation. The user must specify \textit{samp.prior.beta}, a matrix of samples for $\beta$, and \textit{samp.prior.lambda}, a list of matrices where each matrix represents the sampling prior for the baseline hazards for each stratum. The number of columns of each matrix must be equal to the number of intervals for that stratum. 

Now we describe strategies to elicit the sampling priors, as detailed in \cite{Psioda_Ibrahim_2019}. Suppose one wants to test the hypotheses $$H_0: \beta_1 \ge 0$$ and $$H_1: \beta_1 < 0.$$ To elicit the sampling prior for $\beta_1$ to compute power, one can simply sample from a truncated normal distribution with negative mean, so that the mass of the prior falls in the alternative space. Conversely, to compute the type I error rate, one can  sample from a truncated normal distribution with positive mean, so that the mass of the prior falls in the null space. Next, to generate the sampling prior for the other parameters, $\beta_2, \dots, \beta_P$ and $\lambda$ (the same sampling prior is assumed for $\lambda_0$), one can use the posterior samples given the historical data as the discrete approximation to the sampling prior. The function \textit{phm.fixed.a0()} generates such posterior samples if the \textit{current.data} argument is set to FALSE and $a_{0j}=1$ for $j=1,\dots,J$. This method is illustrated in the case study in the following section. 

\section{Case Study: Melanoma Clinical Trial Design}\label{melanoma}

We consider the high-risk melanoma trial design application in \cite{Psioda_Ibrahim_2019}, and demonstrate how \textbf{BayesPPDSurv} can be used for coefficient estimation as well as power and type I error rate calculations for time-to-event data in Bayesian clinical trial designs that incorporate historical information. 

Interferon Alpha-2b (IFN) is an adjuvant chemotherapy for deep primary or regionally metastatic melanoma. The E1684 trial \citep{kirkwood_1996} and the E1690 trial \citep{kirwood_2000} were randomized controlled trials conducted to evaluate the efficacy of IFN for melanoma following surgery. The studies classified the subjects into four disease stage groups. Following \cite{Psioda_Ibrahim_2019}, we restrict our attention to patients in disease stage four, i.e., regional lymph node recurrence at any interval after appropriate surgery for primary melanoma of any depth. The primary outcome is relapse-free survival. The number of positive lymph nodes at lymphadenectomy is used as a stratification variable ($\le 2$ vs. $\ge 3$) due to its prognostic value. We compare patients who received the IFN treatment to those who received observation (OBS). Table \ref{summary} summarizes the total number of events and the total risk time by treatment group and number of positive lymph nodes for the two studies.

\begin{table}[H]
\centering
\setlength{\extrarowheight}{5pt}
\begin{tabular}{cccccc}
\hline
Study & Treatment & \# Nodes & Sample Size & \# Events & Risk Time \\[8pt]
\hline
E1684 & OBS & $\le 2$ & 37 & 26 & 88.4\\
   & & $\ge 3$ & 47 & 36 & 105.8\\
 &IFN & $\le 2$ & 44 & 21 & 176.3\\
&    & $\ge 3$ & 39 & 31 & 81.1\\
\hline
E1690 & OBS & $\le 2$ & 51 & 23 & 122.4\\
   & & $\ge 3$ & 53 & 42 & 78.9\\
 &IFN & $\le 2$ & 51 & 29 & 123.1\\
&    & $\ge 3$ & 59 & 36 & 137.0\\
\hline
\end{tabular}
\caption{Relapse-free survival data by treatment group and number of positive nodes for trials E1684 and E1690. The number of positive lymph nodes is used as a stratification variable.}
\label{summary}
\end{table}

First, suppose we are interested in analyzing the relationship between relapse-free survival and the IFN treatment for the E1690 study after incorporating historical information from the E1684 study. We build a PWCH-PH model with two strata, stratum-specific baseline hazard parameters (unshared between current and historical datasets) and one covariate (the treatment indicator). For stratum 1 (\# nodes $\le 2$), we use four time intervals (selected using the deviance information criterion (DIC)). The baseline hazard parameters are $\lambda_{1,1}$, $\lambda_{1,2}$, $\lambda_{1,3}$, $\lambda_{1,4}$. For stratum 2 (\# nodes $\ge 3$), we use three time intervals (also selected using DIC). The baseline hazard parameters are $\lambda_{2,1}$, $\lambda_{2,2}$, $\lambda_{2,3}$. We use the default change points, which are determined so that an approximately equal number of events are observed in each time interval for the pooled current and historical datasets. We use the default initial prior for $\beta$, a normal prior with mean zero and variance $10^3$. We use the default priors for $\lambda$ and $\lambda_0$, which are independent non-informative gamma priors with shape and rate parameters equal to $10^{-5}$. The code below demonstrates the analysis with the E1684 study data as prior information incorporated through a
power prior with $a_0$ fixed at 0.5.

\begin{verbatim}
data(melanoma)
hist <- melanoma[melanoma$study=="1684",]
current <- melanoma[melanoma$study=="1690",]
n.intervals <- c(4,3)
nMC <- 10000
nBI <- 200
historical <- list(list(time=hist$failtime, event=hist$rfscens, 
                        X=as.matrix(hist[,"trt"]), S=hist$stratum))
set.seed(1)
result <- phm.fixed.a0(time=current$failtime, event=current$rfscens, 
                       X=as.matrix(current[,"trt"]), S=current$stratum, 
                       historical=historical, a0=0.5, n.intervals=n.intervals, 
                       nMC=nMC, nBI=nBI)
> quantile(result$beta_samples)
        0%        25%        50%        75%       100% 
-0.7950745 -0.3694794 -0.2727719 -0.1737096  0.2939300 
> colMeans(result$lambda_samples[[1]])
[1] 0.4594304 0.5089287 0.3052143 0.1188887
> colMeans(result$lambda_samples[[2]])
[1] 1.0761066 0.7789043 0.2019115
> colMeans(result$lambda0_samples[[1]])
[1] 0.67388605 0.75062812 0.35459671 0.06557007
> colMeans(result$lambda0_samples[[2]])
[1] 1.3381514 0.7562455 0.1853991

\end{verbatim}

Table \ref{coefficients} displays the posterior mean, standard deviation and 95\% credible interval for $\beta$ and elements of $\lambda$. There is weak evidence
suggesting a negative association between IFN and time-to-relapse.

\begin{table}[H]
\centering
\begin{tabular}{cccc}
\hline
Parameter & Mean & SD  & 95\% CI \\
\hline
$\beta$ & -0.27 & 0.15 & (-0.57, 0.01)\\
$\lambda_{1,1}$ & 0.46 & 0.14 & (0.23, 0.77) \\
$\lambda_{1,2}$ & 0.51 & 0.15 & (0.26, 0.83) \\
$\lambda_{1,3}$ & 0.31 & 0.09 & (0.16, 0.50)\\
$\lambda_{1,4}$ & 0.12 & 0.03 & (0.07, 0.19) \\
$\lambda_{2,1}$ & 1.08 & 0.23 & (0.68, 1.57) \\
$\lambda_{2,2}$ & 0.78 & 0.16 & (0.50, 1.12) \\
$\lambda_{2,3}$ & 0.20 & 0.04 & (0.12, 0.30) \\
\hline
\end{tabular}
\caption{Posterior mean, standard deviation, and 95\% credible interval for $\beta$ and elements of $\lambda$ for the E1690 study incorporating historical data E1684 via a power prior with $a_0=0.5$. The parameter $\lambda_{s,k}$ represents baseline hazard for time interval $k$ for stratum $s$.}
\label{coefficients}
\end{table}

Next, our goal is to design a new trial incorporating the E1684 study using the power prior and the normalized power prior. We first specify the characteristics of trial data simulation. Let $\nu$ be the number of events at which the trial will stop and let $n$ be the total number of subjects enrolled. For each $\nu$, we take $n = 3\nu$. We assume a subject's enrollment time follows a uniform distribution over a 4-year period. We allocate 50\% of the subjects to the treatment group. For stratum allocation, we sample from the stratum indices of the historical data with replacement. We assume there is only administrative censoring which occurs when $\nu$ events have accrued.  In the data generation phase, we use the default change points which are determined so that an approximately equal number of events are observed in each time interval for the historical dataset. When analyzing the generated data, we use the default change points which are determined so that an approximately equal number of events are observed in each time interval for the pooled current and historical datasets. The baseline hazard parameters are not shared between the current and historical data. The same set of default priors are used for $\beta$, $\lambda$ and $\lambda_0$ as before.

 We compute the Bayesian power and type I error rate for a few sample sizes for tests of the hypotheses $$H_0: \beta \ge 0$$ and $$H_1: \beta < 0.$$ We use two sets of sampling priors, the default null (DN) and default alternative (DA) sampling priors, and the frequentist null (FN) and point-mass alternative (PA) sampling priors. The default null sampling prior is defined as $\pi_0^{(s)}=\pi(\theta|D_0, \beta \ge 0)$. The default alternative sampling prior is defined as $\pi_0^{(s)}=\pi(\theta|D_0, \beta < 0)$. The FN sampling prior is a point-mass prior at $\beta=0$, which results in a Bayesian type I error rate that is closely related to the standard frequentist type I error rate. The PA sampling prior is a point-mass prior centered on $\beta=-0.27$, the posterior mean of $\beta$ for the E1684 dataset, which results in Bayesian power that is closed related to the standard frequentist notion of power (i.e., power at a point). The sampling prior for $\lambda$ is a point-mass prior centered on the posterior mean of $\lambda$ for the E1684 dataset. We first obtain the posterior samples of $\beta$ and $\lambda$ using only the historical data, and then subset to the posterior samples of $\beta$ that are greater than zero for the DN sampling prior and to the samples that are less than zero for the DA sampling prior. The DN and DA sampling priors can be easily acquired with the following code with \textit{current.data=FALSE} and \textit{a0=1}. Here, \textit{nMC} is set to 10,000, but depending on the dataset, \textit{nMC} may need to be larger to ensure enough entries in the truncated sampling priors.

 \begin{verbatim}
nMC <- 10000
nBI <- 200
set.seed(1)
samples <- phm.fixed.a0(historical=historical, a0=1, n.intervals=n.intervals, 
                       current.data=FALSE, nMC=nMC, nBI=nBI)
beta_priors <- samples$beta_samples
DN_beta_samp_prior <- as.matrix(beta_priors[beta_priors[,1] > 0, ])
DA_beta_samp_prior <- as.matrix(beta_priors[beta_priors[,1] < 0, ])
lambda_samp_prior <- samples$lambda_samples
\end{verbatim}
Now we are ready compute the power and type I error rate using the power prior via the function \textit{power.phm.fixed.a0()}. The following code computes the power for $\nu=350$ and $a_0=0.6$ using the default alternative sampling prior.

\begin{verbatim}
set.seed(1)
a0 <- 0.6
n.events <- 350
n.subjects <- n.events * 3
nMC <- 10000
nBI <- 200
N <- 10000
result <- power.phm.fixed.a0(historical=historical, a0=a0, n.subjects=n.subjects, 
            n.events=n.events, n.intervals=n.intervals, 
            samp.prior.beta=DA_beta_samp_prior, samp.prior.lambda=lambda_samp_prior, 
            dist.enroll="Uniform", param.enroll=4, nMC=nMC, nBI=nBI, 
            delta=0, nullspace.ineq=">", N=N)
\end{verbatim}
Table \ref{powers} displays the power and type I error rates for $\nu=350$ and $\nu=710$ for $a_0$ values of $0$, $0.2$ and $0.6$ using the default sampling priors and the point-mass sampling priors. We obtain 50,000 posterior samples using the package's custom slice sampler after 200 burn-ins for each of the 10,000 simulated datasets. The results in Table \ref{powers} are comparable to the results in  Figure 2 in \cite{Psioda_Ibrahim_2019} (i.e., the same up to Monte Carlo error). We observe that the power increases with $a_0$ and sample size as expected. We can also see that the default sampling priors yield average rates that are often lower than rates based on point-mass priors.

\begin{table}[H]
\centering
\setlength{\extrarowheight}{5pt}
\begin{tabular}{ccccc}
\hline
\thead{Null sampling prior/\\Alt sampling prior} & $a_0$ & Number of events & Type I error rate & Power \\[8pt]
\hline
DN/DA & 0 & 350 & 0.0121 & 0.3947 \\ 
      &   & 710 & 0.0094 & 0.5868 \\ 
	  & 0.2 & 350 & 0.0134 & 0.4280\\
	  &     & 710 & 0.0113 & 0.6055\\
	  & 0.6 & 350 & 0.0242 & 0.4917\\
	  &     & 710 & 0.0154 & 0.6504\\
\hline
FN/PA & 0 & 350 & 0.0239 & 0.3145  \\ 
      &   & 710   & 0.0226 & 0.5682 \\ 
	  & 0.2 & 350 & 0.0298& 0.3545\\
	  &     & 710 & 0.0349& 0.5983\\
	  & 0.6 & 350 & 0.0508& 0.4435\\
	  &     & 710 & 0.0474& 0.6851\\
\hline
\end{tabular}
\caption{Power and type I error rates for $\nu=350$ and $\nu=710$ for $a_0$ values of $0$, $0.2$ and $0.6$ using the default sampling priors and the point-mass sampling priors. We run 50,000 iterations of the slice sampler with 200 burn-ins for each of the 10,000 simulated datasets.}
\label{powers}
\end{table}

Finally, we demonstrate how to estimate operating characteristics for a new trial incorporating the E1684 study with the normalized power prior where $a_0$ is modeled as random. We compute power and type I error with the DA and DN sampling priors using the function \textit{power.phm.random.a0()}.
By default, the function \textit{power.phm.random.a0()} approximates the normalized power prior for $\beta$, $\pi(\beta|D_0)$, with a single multivariate normal distribution. We now show how to approximate $\pi(\beta|D_0)$ with a mixture of multivariate normal distributions. We first use the \textit{approximate.prior.beta()} function to acquire discrete samples of $\beta$ from the normalized power prior, and then use the R package \textbf{mixtools} \citep{mixtools} to compute a mixture of multivariate normal distributions that best approximates the samples. In the code below, we put the output of the function \textit{normalmixEM}, a weighted mixture of two normal distributions, into a list, which then becomes the input to the parameter \textit{prior.beta.mvn}. 

\begin{verbatim}
set.seed(1)
nMC <- 10000 
nBI <- 200
N <- 10000
prior.beta <- approximate.prior.beta(historical, n.intervals, 
                   prior.a0.shape1=1, prior.a0.shape2=1, 
                   nMC=nMC, nBI=nBI)
library(mixtools)
mix <- normalmixEM(prior.beta)  
list_mixture <- list(list(mix$mu[1], as.matrix(mix$sigma[1]), mix$lambda[1]),
                     list(mix$mu[2], as.matrix(mix$sigma[2]), mix$lambda[2]))
result <- power.phm.random.a0(historical=historical, n.subjects=n.subjects, 
            n.events=n.events, n.intervals=n.intervals, 
            prior.beta.mvn=list_mixture,
            samp.prior.beta=DA_beta_samp_prior, samp.prior.lambda=lambda_samp_prior,  
            dist.enroll="Uniform", param.enroll=4, nMC=nMC, nBI=nBI, 
            delta=0, nullspace.ineq=">", N=N)
\end{verbatim}

In Table \ref{power_npp}, we compare the power and type I error rate calculations using the normalized power prior with a beta($10^3$, $10^3$) prior for $a_0$, with a power prior with $a_0=0.5$. We expect the results to be quite similar since the prior on $a_0$ for the normalized power prior reflects a high degree of certainty that the correct value is near 0.5. The DN/DA sampling priors are used, and we run 10,000 iterations of the slice sampler with 200 burn-ins for each of the 10,000 simulated datasets. The results are quite similar, indicating that our algorithm for approximating the normalized power prior provides reliable estimates. 

\begin{table}[H]
\centering
\setlength{\extrarowheight}{5pt}
\begin{tabular}{llcc}
\hline
Model & $a_0$ & Type I error rate & Power \\[8pt]
\hline
power prior & 0.5 & 0.029 & 0.770 \\ 
normalized power prior & beta($10^3$, $10^3$)& 0.028 & 0.768\\ 
\hline
\end{tabular}
\caption{Power and type I error rate computed with power.phm.fixed.a0() with $a_0$ fixed at $0.5$, and power.phm.random.a0() with a beta($10^3$, $10^3$) prior on $a_0$. The DN sampling prior is used to compute the type I error rate and the DA sampling prior is used to compute power. The number of events used is $350$. We run 10,000 iterations of the slice sampler with 200 burn-ins for each of the 10,000 simulated datasets.}
\label{power_npp}
\end{table}

\section{Discussion} \label{sec:discussion}

\textbf{BayesPPDSurv} facilitates Bayesian power and type I error rate calculations using the power and normalized power prior for time-to-event outcomes using a PWCH-PH model. We implement a flexible stratified version of the model, where the historical data can be used to inform the treatment effect, the effect of other covariates in the regression model, as well as the baseline hazard parameters. We develop a novel algorithm for approximating the normalized power prior that eliminates the need to compute the normalizing constant. The package also has features that semi-automatically generate the sampling priors from the historical data. 

Future versions of the package will accommodate cure rate models. Another possible feature is the computation of optimal hyperparameters for the beta prior on $a_0$ to ensure that the normalized power prior adapts in a desirable way to prior-data conflict or prior-data agreement, based on the work of \cite{shen_2023}. 

\bibliographystyle{chicago}
\bibliography{shen}

\end{document}